%% file: inhomo.tex
\documentclass[iop]{emulateapj}
\usepackage{natbib}
\usepackage{graphicx}
\usepackage{rotating}

\newcommand{\el}{$\ell$}
\newcommand{\m}{$m$}
\newcommand{\lm}{$(\ell$, $m)$}
\newcommand{\mone}{$(1$, $\mbox{-}1)$}
\newcommand{\monep}{$(1$,~$1)$}
\newcommand{\mzero}{$(1$,~$0)$}
\newcommand{\mpm}{$(1$,~$\pm1)$}
\newcommand{\kms}{km\,s$^{-1}$}

\newcommand{\rc}{$R_{\rm c}$}
\newcommand{\ica}{$I_{\rm Ca}$}

\begin{document}

\title{Pulsational Mapping of Calcium Across the Surface of a White Dwarf}   
\author{Susan E. Thompson\altaffilmark{1,2}, M. H. Montgomery\altaffilmark{2,3}, T. von Hippel\altaffilmark{3,4}, A. Nitta\altaffilmark{5}, J. Dalessio\altaffilmark{1,2}, J. Provencal\altaffilmark{1,2}, W. Strickland\altaffilmark{6}, J. A. Holtzman\altaffilmark{7}, A. Mukadam\altaffilmark{8}, D. Sullivan\altaffilmark{9}, T. Nagel\altaffilmark{9}, D. Koziel-Wierzbowska\altaffilmark{10}, S. Zola\altaffilmark{10,11}, T. Kundera\altaffilmark{10}, E. Kuligowska\altaffilmark{10}, M. Winiarski\altaffilmark{11},  M. Drozdz\altaffilmark{11}, W. Ogloza\altaffilmark{11}, Zs. Bogn\'ar\altaffilmark{12}, G. Handler\altaffilmark{13},  A. Kanaan\altaffilmark{14}, T. Ribeira\altaffilmark{14}, R. Rosen\altaffilmark{15}, D. Reichart\altaffilmark{16}, J. Haislip\altaffilmark{16}, B. N. Barlow\altaffilmark{16},  B. H. Dunlap\altaffilmark{16}, K. Ivarsen\altaffilmark{16}, A. LaCluyze\altaffilmark{16}, \& F. Mullally\altaffilmark{17}
}

\altaffiltext{1}{Department of Physics and Astronomy, University of Delaware, Newark, DE 19716, USA; sthomp@physics.udel.edu}
\altaffiltext{2}{Delaware Asteroseismic Research Center, Mount Cuba Observatory, Greenville, DE 19807, USA}
\altaffiltext{3}{Department of Astronomy, University of Texas at Austin, 1
University Station C1400, Austin, TX 78712-0259, USA}
\altaffiltext{4}{Department of Physics and Astronomy, Siena College, 515 Loudon Road, Loudonville, NY 12211, USA}
\altaffiltext{5}{Gemini Observatory, 670 N A'ohoku Pl., Hilo, HI 96720, USA ; Subaru Telescope, 650 N A'ohoku Pl., Hilo, HI 96720, USA}
\altaffiltext{6}{P. J. Meyer Observatory, Clifton, TX 76634 USA}
\altaffiltext{7}{Department of Astronomy, New Mexico State University, Box 30001, 1320 Frenger St., Las Cruces, NM 88003, USA}
\altaffiltext{8}{Department of Astronomy, University of Washington, Seattle, WA 98195-1580, USA}
\altaffiltext{9}{School of Chemical \& Physical Sciences, Victoria University of Wellington, PO Box 600, Wellington, New Zealand}
\altaffiltext{10}{Astronomical Observatory, Jagiellonian University, ul. Orla 171, 30-244 Cracow, Poland}
\altaffiltext{11}{Mt. Suhora Observatory, Pedagogical University, ul. Podchorazych 2,30-084 Krakow, Poland}
\altaffiltext{12}{Konkoly Observatory, P.O. Box 67, H-1525 Budapest XII, Hungary}
\altaffiltext{13}{Institut f\"ur Astronomie, Universit\"at Wien,
T\"urkenschanzstrasse 17, A-1180 Wien, Austria}
\altaffiltext{14}{Departamento de F\'isica Universidade Federal de Santa Catarina, CP 476, 88040-900 Florianópolis, SC, Brazil }
\altaffiltext{15}{National Radio Astronomy Observatory, Green Bank, WV 24944, USA}
\altaffiltext{16}{Department of Physics and Astronomy, University of North Carolina, Chapel Hill, NC 27599 USA}
\altaffiltext{17}{Department of Astrophysical Sciences, Princeton University, Princeton, NJ 08544, USA}  
  

\begin{abstract} 
We constrain the distribution of calcium across the surface of the white dwarf star G29-38 by combining time series spectroscopy from Gemini-North with global time series photometry from the Whole Earth Telescope.  G29-38 is actively accreting metals from a known debris disk. Since the metals sink significantly faster than they mix across the surface, any inhomogeneity in the accretion process will appear as an inhomogeneity of the metals on the surface of the star. We measure the flux amplitudes and the calcium equivalent width amplitudes for two large pulsations excited on G29-38 in 2008.  The ratio of these amplitudes best fits a model for polar accretion of calcium and rules out equatorial accretion. 

\end{abstract}

\section{Introduction}

The well known G29-38 is a bright white dwarf star with a H atmosphere (DA) that falls in the instability strip (DAV), has metals apparent in its photosphere (DAVZ), and has a known debris disk (DAVZd). The combination of these attributes affords us the opportunity to measure the geometry of accretion from the debris disk onto the white dwarf star.

As a DAV, G29-38 shows multi-periodic, large amplitude pulsations ($>2$\%) with periods between 200 and 1000\,s.  These are non-radial, g-mode pulsations that likely arise from convective driving at the surface of the star \citep{brickhill92,gw1}.  Each pulsation is a standing wave that can be described by three eigenvalues, $k$, \el, and \m. The first describes the radial nodes while the last two, \lm, denote the spherical harmonic, or the surface geometry of the pulsation.  G29-38 does not always excite the same modes, selecting several modes from the set of possible eigenvalues \citep{kleinman98}. Matching the observed periods with the correct eigenvalues is an important step in order to use pulsators like G29-38 to measure stellar interiors with a technique known as asteroseismology. As a bright DAV, G29-38 has been the target for developing techniques to measure \lm\ of each excited mode. Techniques such as time series spectroscopy \citep{vk00,c00,thompson08}, combination modes \citep{wu01}, and light curve fitting \citep{montgomery05} have found excited \el=2 modes among primarily \el=1 modes.

G29-38 became part of a larger mystery when the metals Ca, Mg and Fe were found in its photosphere \citep{koester97}.  White dwarf stars have very large gravities, and the diffusion timescale for Ca to precipitate out of the photospheres for those with temperatures less than $\sim20,000$\,K is much less than their cooling times \citep{koester09, koester06}.  For G29-38 this diffusion timescale is less than 2 weeks, meaning G29-38 must be actively accreting metals. The source of the metals, at least for G29-38, has been identified with the discovery of a surrounding debris disk \citep{tokunaga90, graham90, reach05}.  The infrared excess is well modeled by sub-micron silicates and amorphous carbon lying between 1-5\,R$_\sun$ from the star and is possibly formed from disrupted minor bodies \citep{farihi09,jura08}. The material accretes onto G29-38, producing the metal lines seen in its photosphere. 

The accretion process onto G29-38 does not need to be geometrically uniform.  For instance, the dust around G29-38 could be a thin disk that accretes material near the stellar equator. Alternatively, a magnetic field may guide material onto the star's poles. Once the metals are accreted they begin to sink out of the photosphere. At the same time the metals are mixed across the stellar surface by the turbulent viscosity produced by the convection zone. As long as the timescale for gravitational settling is much shorter than the mixing timescale, the inhomogeneity of metals on the surface of the star will be maintained. \citet{montgomery08} estimate the gravitational settling to be 30-50 times shorter than the horizontal mixing. If the accretion is indeed non-uniform, then the surface of G29-38 will have a non-uniform distribution of Ca across its surface. 

\citet{montgomery08} proposed a technique to use the pulsations of G29-38 to measure any inhomogeneity of the surface metals. This technique is based on the fact that the strength of the metal lines is a function of temperature. As the pulsations cause temperature variations on the surface of the star, the equivalent width (EW) of any metal line will vary along with the pulsations. For G29-38, the size of this effect is as large as 10\% for a single pulsation mode \citep{vonhippel07}. The pulsation temperature variations are described by a spherical harmonic, and since g-modes all have \el$\ge$1, the temperature variations are not uniform across the surface.  As a result, the amplitude of the EW variations relative to the photometric amplitude ($I\equiv A_{\rm EW}/A_{\rm flux}$) will depend on the spherical harmonic of the pulsation mode and on the distribution of the metals.  When the pulsation geometry aligns with the metal inhomogeneity, $A_{\rm EW}$ will be larger. For example, if the metals followed an equatorial distribution, the EW variation due to the pulsations would be larger for a \monep\ mode that primarily pulsates at the equator than a \mzero\ mode that primarily pulsates at the poles. 

The goal of this paper is to measure the geometry of Ca across G29-38 using the technique of \citet{montgomery08}.  We obtained time series spectroscopy from Gemini-North in order to measure the EW variations of the Ca K 3933\AA\ line. We also obtained time series photometry from a collection of telescopes associated with the Whole Earth Telescope to accurately measure the photometric amplitudes of the pulsation modes. G29-38 excited two large amplitude triplets during our 2008 data campaign, each split by 15\,$\mu$Hz. We use the two largest modes, with different values of \m, to constrain the inhomogeneity of Ca on the surface of G29-38. Finally, we apply this same analysis to previous time series spectroscopy of G29-38: the VLT (Very Large Telescope) data published by \citet{thompson08} and the Keck Observatory data published by \citet{vk00}.

\section{Data}
We received 9.5\,hr of time series spectroscopy of G29-38 from GMOS \citep[Gemini Multi-Object Spectrograph,][]{gmos} on Gemini-North in 2008. Concurrently we used the Whole Earth Telescope  \citep{wet} to obtain over 300\,hr of time series photometry from 10 different telescopes.  The third priority (``band-3''), queue-scheduled Gemini observations were obtained during two nights in October and three nights in November.  These observations were sufficiently separated that we treated the data as two separate measurements of G29-38's pulsations. Luckily, the same large pulsation modes were present during both months, allowing us to easily average our measurements.

\subsection{Photometry}
We obtained the time series photometry for two distinct purposes. First, we needed enough data to clearly identify the pulsation frequencies present in G29-38. Second, we needed to correctly measure the flux amplitude of each pulsation. We obtained photometry of G29-38 from early August through early December in 2008. The majority of the coverage is in September, while the Gemini spectra were not taken until October and November.  Several observers at small telescopes were on-call to take these observations when Gemini observed. The observations we present here are split into three groups. The first, in September, is the most extensive but also lacks corresponding Gemini observations. The second, in October, and the third, in November, each have $\sim$5\,hr of accompanying time series spectra. See Tables~1-3 for a list of the photometric observations. 

In September we obtained 29 separate runs from 8 different telescopes, totaling more than 177\,hr of photometry.  In October we obtained 13 separate runs from 5 different telescopes, totaling more than 50\,hr of photometry. In November, we obtained 15 runs from 6 different telescopes, totaling more than 63\,hr of photometry. Each observatory used 5-15\,s exposure times with a minimal readout time of less than 15\,s.  We used a broadband, red cutoff filter, \emph {BG40} or similar, to reduce the variations between the sensitivities of each telescope and CCD camera.  The PROMPT telescope (Panchromatic Robotic Optical Monitoring and Polarimetry Telescopes), lacking a \emph{BG40}, used a Sloan $g$ filter. Though the $g$ filter lets through less red light, this has little effect on the measured amplitudes of the DAV pulsation modes ($\la$3\%). 

To reduce the photometric images, we applied dark, bias and flat-field corrections to our image frames. We then performed aperture photometry using IRAF's {\sc PHOT} routine on each image. We used the aperture that gave the highest signal-to-noise for each night of data. To create the final light curve, we divided by a bright star, removed bad points, and corrected for differential extinction using a program called WQED \citep{wqed}. Finally we applied a barycentric correction to our times using the method of \citet{Stumpff80}.

\subsection{Spectra}

\begin{figure*}[ht]
\centering
\includegraphics[scale=.5, angle=270]{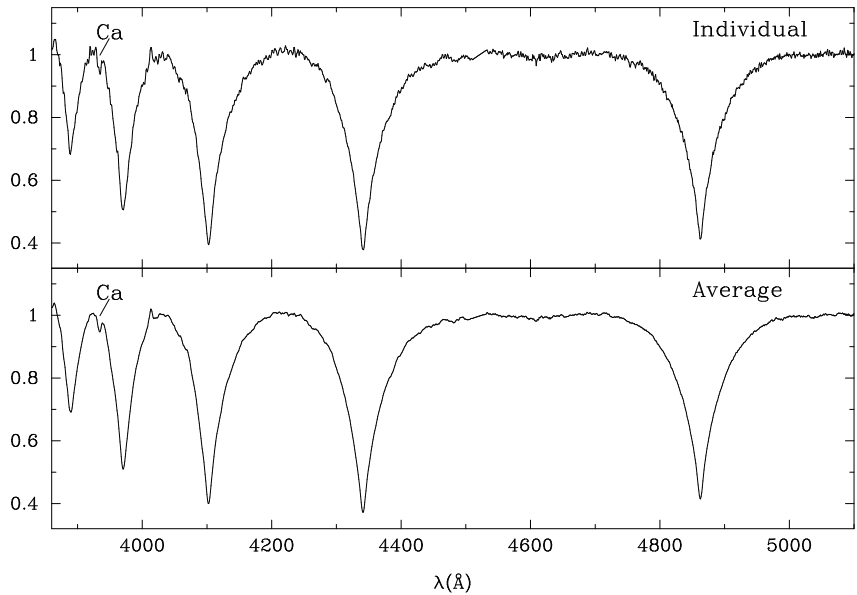}
\caption{\label{f:spec}A portion of the individual and average spectrum of G29-38 taken with GMOS on Gemini-North. Each spectrum has been normalized with a low-order spline function for presentation purposes. The Ca K (3933\,\AA) line has been labeled.}
\end{figure*}

We obtained a total of 9.5\,hr of time-series spectroscopy on 5 different nights in 2008 (October 16, October 18, November 03, November 04, November 06), approximately half of our allocated queue-scheduled Gemini observations.  Each spectrum has an exposure time of 55\,s and a readout time of 13\,s, yielding a Nyquist period of 136\,s. We reduce these spectra from Gemini IRAF GMOS routines, applying flats, biases and arc-lamps taken at the beginning of the night.  The resulting dispersion is 0.9\,\AA\ while the resolution is $\le$3.2\,\AA, set by a 1.5'' slit.  The signal-to-noise of the spectra is typically 80, while the average spectrum for one night has a signal-to-noise of $\sim300$ (see Figure~\ref{f:spec}). The spectra cover the optical range from 3700-6400\,\AA, including both Mg {\sc II} (4480\,\AA) and Ca K (3933\,\AA). While both are present in the spectra, the Mg line is much weaker, and we are unable to measure variations in its line strength.

We use the spectra primarily to measure the variations in the EW of the Ca line. However the prominent Balmer lines also change with the pulsations.  We use the H lines to measure the line shape variations to identify the spherical degree, \el, of each mode (see \S\ref{s:shape}).

\section{Measuring the Pulsations}

\subsection{Flux Variations}
\begin{figure*}[ht]
\centering
\includegraphics[scale=.5, angle=270]{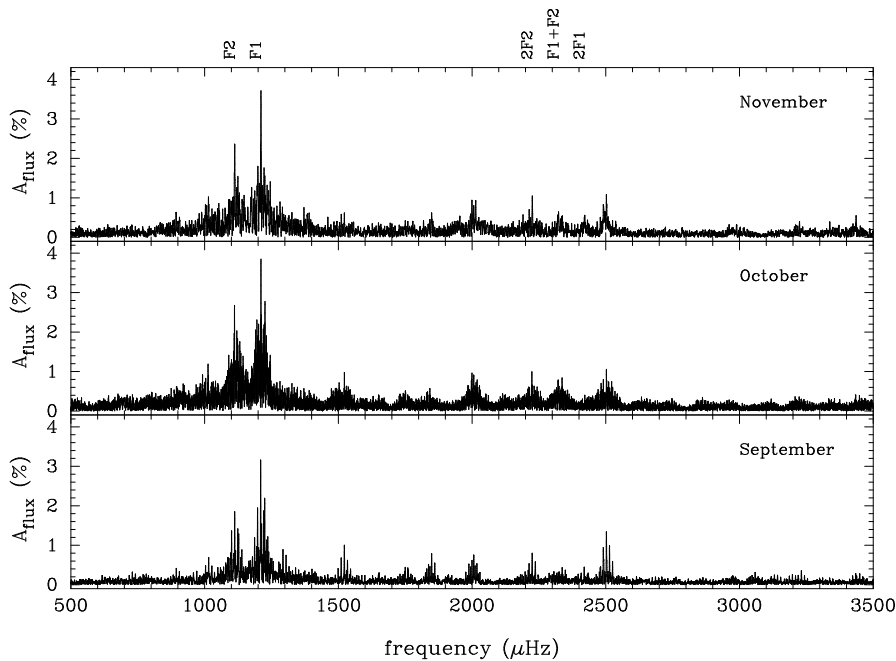}
\caption{\label{f:ftphot} The Fourier transforms of light curves measured during the three G29-38 photometry campaigns. The frequencies of F1, F2 and their combinations are denoted at the top of the plot.}
\end{figure*}

\begin{figure}[ht]
\centering
\includegraphics[scale=.31, angle=270]{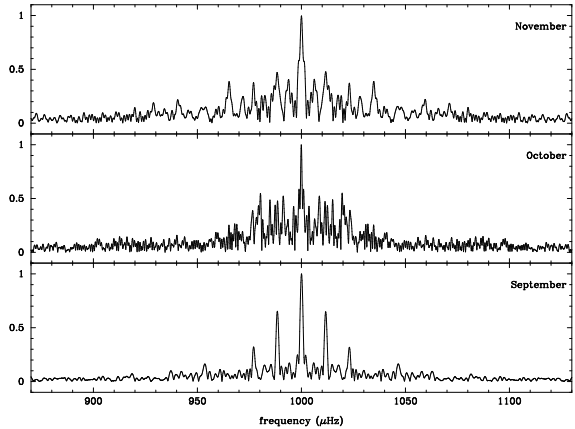}
\caption{\label{f:ftwin}The Fourier window for each month of data is shown at 1000$\mu$Hz. September had the most coverage and thus the fewest alias peaks.}
\end{figure}

We perform a Fourier transform (FT) of the light curves from the September, October and November data sets (see Figure~\ref{f:ftphot}).  To judge the alias peaks created from the gaps in our data, we present the Fourier windows of each data set in Figure~\ref{f:ftwin}. G29-38 appears to have excited essentially the same modes during all our observations.  

We fit and remove the largest peaks found in the FT using the program PERIOD04 \citep{p04}.  While each group of data is dominated by the same sets of modes, their frequencies appear to wander from month to month. \citet{provencal09} have observed similar behavior in other pulsating white dwarf stars. We use the September photometry as a guide to select the dominant modes and their combinations. We identify modes from subsequent months with the same name (e.g., F1) if they lie within 3\,$\mu$Hz. In the rare case that a statistically significant mode does not exist in the FT from that month, we perform a nonlinear fit at the largest peak near the September frequency (F4 in November is only marginally significant, but was fit for completeness). While significant peaks may remain, we do not discuss them here. We use only the largest modes and their combinations to determine the metal distribution across G29-38. 

We provide the resulting list of frequencies, amplitudes and phases from each data set in Tables~4-6. The zero time for all phases is BJED=2454700.0. We calculate errors for the photometry with a 100 point Monte Carlo simulation using the program PERIOD04. We also fit the combination and harmonic peaks present in the FT, by constraining each frequency to the exact frequency derived by the sum of the parent mode(s).

\subsection{Equivalent Width Variations}
\label{s:ew}
The convection zone of a DAV extends into its photosphere, so the time
series analysis of its spectral lines is simpler than for other
pulsating stars. The observed pulsations are g-modes that are
non-propagating (``evanescent'') in the convection zone and the
photosphere. Additionally, the short turn over time ($\la 1$~s) of the
convection zone efficiently mixes the velocity variations, making them
constant as a function of optical depth, and the total (i.e.,
bolometric) flux variations are also constant through these outer
layers \citep[see Figures~2 through 5 of][]{gw2}. From our models we
estimate that these conditions are satisfied from the outer atmosphere
down to optical depths of $\tau \ga 1000$.  In addition, convective
mixing removes any depth dependence of the Ca abundance.  Thus, any
local region of the star's surface can be modeled with a static model
atmosphere having the instantaneous local values of the $T_{\rm eff}$,
$\log g$, and Ca abundance.  This is to be contrasted with the case of
the roAp stars, in which the eigenfunctions can have significant
variations even at small optical depths \citep[e.g., see][]{medupe98}.

To measure the Ca EW variations, we normalize by the continuum and then fit a Gaussian to the Ca line in each spectrum. For low resolution data, simulations show us that fitting a Gaussian more accurately measures the area of the line than does numerical integration across the line.  Since all the spectra were acquired with the same spectrograph and set-up, we averaged the spectra to determine the shape of the continuum. (The continuum in this case is actually the wings of the H$\epsilon$ and H$8$ lines.)  We fit a 5$^{\rm th}$ order polynomial across 35\,\AA\ of the spectrum, after removing the central 5\,\AA\ of the line. This shape is used to flatten the continuum of each spectrum. Prior to measuring the EW, we normalize a second time by a linear trend to remove any remaining brightness and color variations due to the star or the Earth's atmosphere (again, we do not include the central 5\,\AA\ of the line). By fitting the shape of the continuum with the average spectrum we reduce the overall noise that would come from establishing the continuum with much noisier, individual spectra.  The EW is computed as the area removed by the best fit Gaussian divided by the continuum level for each spectrum.  

\begin{figure*}[ht]
\centering
\includegraphics[scale=.5, angle=270]{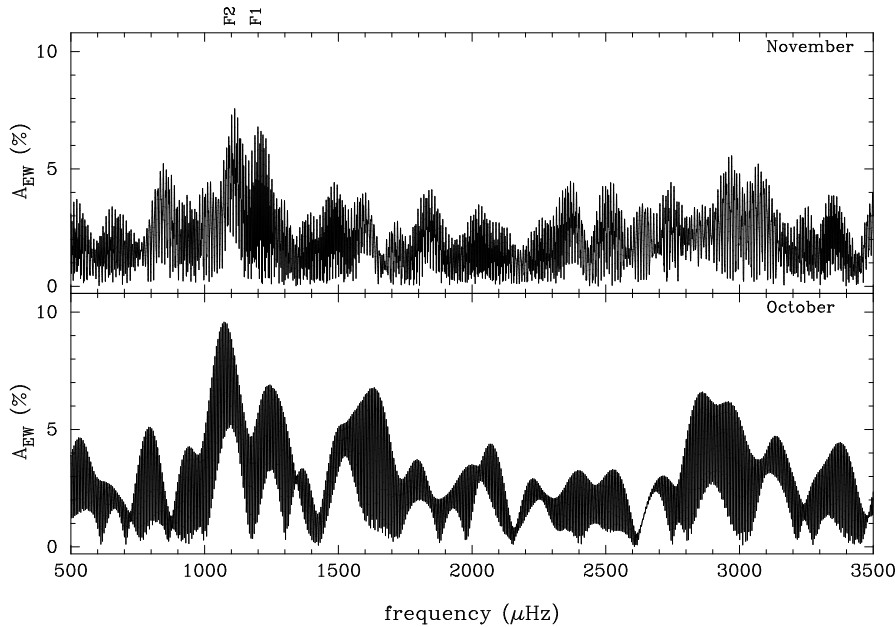}
\caption{\label{f:ew}The Fourier transforms of the fractional Ca EW variations during October and November. F1 and F2 are labeled at the top for reference.}
\end{figure*}

To create fractional variations in EW, we divide by the average EW of the line from each night.  We measure the EW amplitudes by fitting these variations at the exact frequencies measured from the accompanying photometry. In Figure~\ref{f:ew} we present the FT of the EW measurements. The noise level, measured from the average power of the high frequency end of the FT, is 2.8\% in the October data and 2.3\% in the November data. Since we know the frequency from the photometry, the peaks found at the frequencies for F1 and F2 are significant. In both November and October, the probability that F1 in the EW FT is due to noise alone is $1/200$, and similarly the probability that F2 is due to noise alone is less than $1/2000$ \citep{kepler93}. Tables~5 and~6 summarize the amplitudes of the EW variations along with the formal errors for the largest modes present in October and November.  

The above method for measuring the Ca EW suffers from systematic underestimates of the EW because of poor continuum normalization. In the next section we consider how the continuum fitting impacts our EW measurements and we quantify this effect.


\subsection{Systematics in Measuring the Continuum}
\label{s:offset}
The EW amplitudes are measured relative to the average EW. If we fit a continuum that mis-measures the average EW, then the EW amplitude itself will be wrong because we will have falsely added or subtracted area from the line. In general we find that we pick an average EW that is too small because the continuum includes the wings of the broad line, making the assumed continuum across the line have less flux than it should. 

We estimate this offset by fitting model spectra containing the Ca K line \citep{weidemann84}. We create $>$200 model spectra with a dispersion (0.9\,\AA), resolution (3.2\,\AA), and signal-to-noise ($\sim$300), approximately equivalent to our average Gemini spectrum \citep{weidemann84}.  After fitting the continuum and measuring the EW with the same procedure as the data, we find that the measured EW is only $70\pm1$\% of the actual, model EW. If not corrected for this effect, the measured Ca EW amplitude would be too large. 

Our assumption that the continuum shape is identical throughout the pulsation also causes a systematic offset, though a small one.  Because we normalize all spectra by the same continuum, established from the mean spectrum, there is a slight systematic linear trend to the offset. Hotter spectra have weaker Ca lines than the cooler spectra. Additionally, a change in temperature alters the shape of the adjacent H lines and hence alters the effective continuum for the Ca line. Our technique underestimates the Ca EW in the hotter spectra to a larger extent than it does the cooler spectra.  For example, the measured EW is $73$\% of the actual EW for measured EWs that are 10\% larger than the mean. Based on this and other calculations, we correct our measurements using the form $70\%+0.3x$ where $x$ is the per~cent difference between the measured EW and the average EW.  As the pulsations cause EW variations by at most 20\%, this additional linear trend is small, but significant.

Tables~5 and~6 give the EW amplitudes after correcting by these effects. We find the average Ca EW to be $279\pm10$\,m\AA.

\section{Geometry of Pulsation}
To compare our measurements of the ratio between the Ca EW and the flux amplitudes, \ica$=A_{\rm EW}/A_{\rm flux}$ (where we use the subscript to refer to either the element or the mode measured), we must first determine each pulsation mode's surface geometry. Simply put, we need the spherical harmonics, \lm, and the inclination of the pulsation axis relative to our line-of-sight.  Recently, several methods for determining the spherical degree and azimuthal order have been developed. Each method individually leaves room for doubt, but used together we can reach an agreement on the geometry of the pulsation modes. We use the techniques of asteroseismology, combination modes, and line-shape variations to determine the \lm\ of the two largest pulsation modes. The inclination angle is more difficult to determine and is left as a free parameter when we compare our measurements to the inhomogeneity model of \citet{montgomery08}.

\subsection{Multiplet Structure}
\label{s:trip}
\begin{figure*}[ht]
\centering
\includegraphics[scale=.55,angle=270]{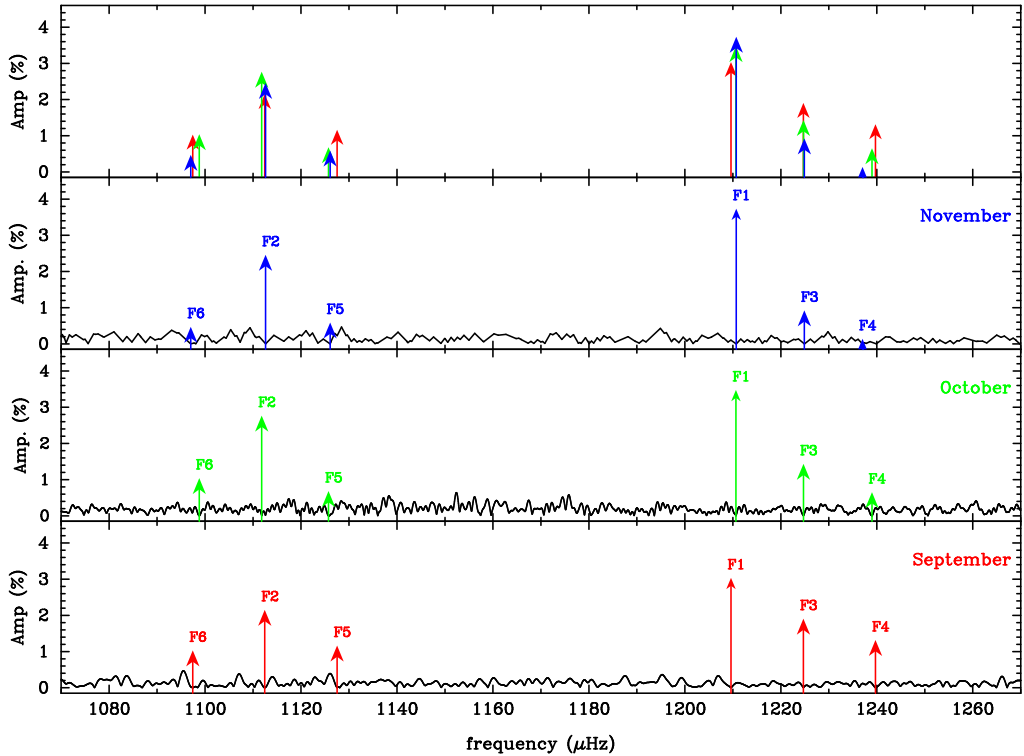}
\caption{\label{f:trip}A schematic illustration of the two 15\,$\mu$Hz triplets observed in 2009 on G29-38, each panel represents a different month of data. Each arrow represents the frequency and the amplitude of a nonlinear fit to the complete lightcurve of each month.  The FT of the residuals is shown in black for each month.}
\end{figure*}

Stars pulsate at frequencies dictated by their interior structure. Asteroseismology, the technique of using the pulsation frequencies to determine the properties of the stellar interiors, predicts that when a star rotates, an $\ell=1$ mode will be split into $2\ell+1$ modes with different values of $m$. The size of the frequency splitting depends on the rate of rotation \citep{winget91}.  Both of our modes are part of clear triplets split by 15 $\mu$Hz, giving us the likely guess that each mode is $\ell$=1.  Figure~\ref{f:trip} shows both triplets, labeled according to Table~4. Unfortunately, white dwarf pulsators do not always excite all members of a multiplet; an $\ell$=2 multiplet may also appear as a triplet leaving some doubt to the identification of our modes.  If both are $\ell=1$ then F1, as the first member of one triplet, and F2, as the central member of another, yield the most likely \lm\ identifications of F1=\mone\ and F2=\mzero.

\subsection{Line shape variations}
\label{s:shape}
Since the extent of limb darkening varies across the broad H features in white dwarf spectra, the shapes of the H lines periodically vary according to the \el\ of the pulsation mode. \citet{robinson95} first introduced the technique to measure how the amplitude of the pulsation varies at different wavelengths for pulsating white dwarf stars. Since then the technique has been successfully applied to bright DAV and DBV stars \citep[e.g.][]{c00,kotak02a,kotak04}.
The line shape variations are independent of the inclination angle and $m$, and can only be used to measure \el.  Recently, \citet{thompson04,thompson08} have shown that fitting a pre-specified functional form to the spectral lines allows this technique to be applied to poorer quality data. Since our observations are of the same star as \citet{thompson08}, with exactly the same technique we can directly compare our results to the previous time series spectroscopy. 

Following the technique described by \citet{thompson08}, we measure spherical degree by fitting the H$\beta$ and H$\gamma$ lines in our Gemini spectra. We describe the variation in the shape of the Balmer lines by fitting the combination of a Gaussian and a Lorentzian to each spectral line. We measure the EW of both the Gaussian and the Lorentzian and measure their variations at the known pulsation periods to determine the fractional amplitudes (denoted $G_{\rm EW}$ and $L_{\rm EW}$, respectively). We then normalize those amplitudes by the flux amplitude of the mode measured by the fitted continuum at the center of the Balmer line (e.g., $A_{\rm 4862}$). By plotting the amplitude of the variations in the normalized areas of the Lorentzian against the Gaussian, we can distinguish between different values of \el. While there is a known offset between the models and the observed modes, different spherical degrees appear to follow the basic trend outlined by the models. 

\begin{figure*}[ht]
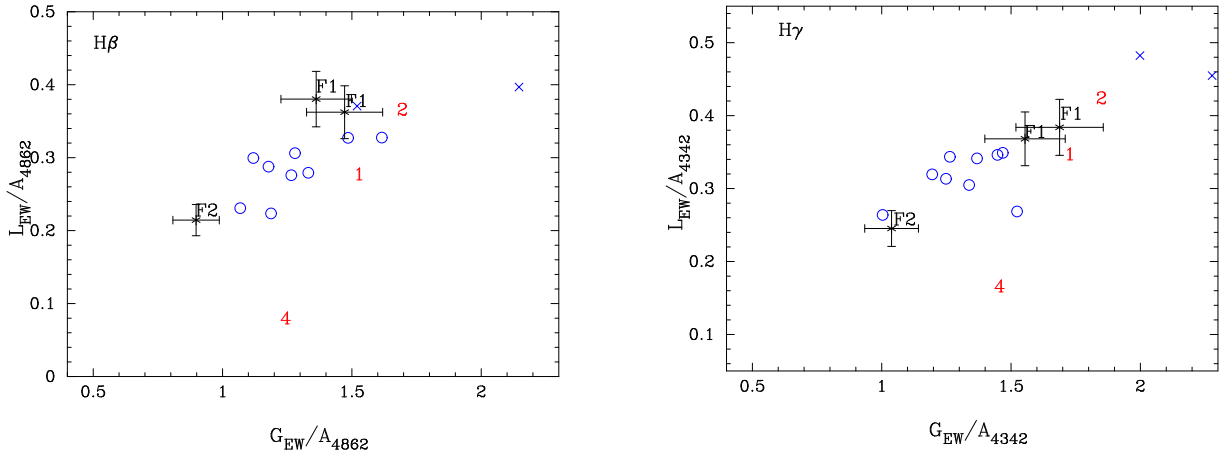

\includegraphics[scale=.35,angle=270]{plots/beta.ps}
\hspace{4em}
\includegraphics[scale=.35,angle=270]{plots/gam.ps}
\caption{\label{f:lineshape}Line shape variations plotted for the H$\beta$ and H$\gamma$ lines on G29-38. The amplitudes of the variations in the Gaussian area are plotted against the Lorentzian area after both are normalized by the amplitude of the mode (A$_{4862}$ and A$_{4342}$ represent the fractional amplitude of the mode measured for the fitted continuum at the center of the line). The integers 1, 2 \& 4 represent the corresponding model \el\ values \citep{weidemann84}. F1 and F2, as measured from our Gemini data, are shown with error bars. The $\ell$=1 modes identified from previous G29-38 spectra are represented as circles \citep{vk00,thompson08}. The previously identified $\ell$=2 modes are presented as {\bf x}'s.}
\end{figure*}

We limit this analysis to the two largest photometric modes (F1 and F2) as they are the only modes large enough to show clear line shape variations.  Figure~\ref{f:lineshape} shows the H$\gamma$ and H$\beta$ variations compared to both a model appropriate for G29-38 and the previous observations of this star \citep{c00,thompson08}. Our Gemini observations were of high enough quality to measure F1 in October and both F1 and F2 in November. The formal error bars from fitting the Gaussian and Lorentzian functions at the known periods are represented in Figure~\ref{f:lineshape}. Both F1 and F2 lie amongst the previously observed $\ell$=1 modes, making it the most likely interpretation of the data. However, given the size of the error bars and the ambiguity in where an $\ell$=1 mode should lie, there is still a nonzero chance that F1 is $\ell$=4 and F2 is $\ell$=2. 

\subsection{Combination modes}
\label{s:comb}
We use combination modes to identify \lm\ of the pulsation modes in the manner described by \citet{wu01}.  The ratio of the combination amplitude ($A_{\rm ij}$) to the parent amplitudes ($A_{\rm i}$ and $A_{\rm j}$) is given by \rc $\equiv$$A_{\rm ij}/n_{\rm ij}A_{\rm i}A_{\rm j}$, where $n_{\rm ij}$ is 1 for harmonics and 2 for combinations. \citet{wu01}, using the surface geometry of each mode and the inclination angle, makes a prediction for this value. Her analysis only uses the first order perturbation, and thus the mode identification technique is most successful when applied to low-amplitude pulsators \citep{handler02,yeates05}. While our F1 and F2 are relatively large amplitude modes, we find that this combination analysis yields consistent results and is still useful for G29-38.

\begin{figure}[ht]
\centering
\includegraphics[scale=.6]{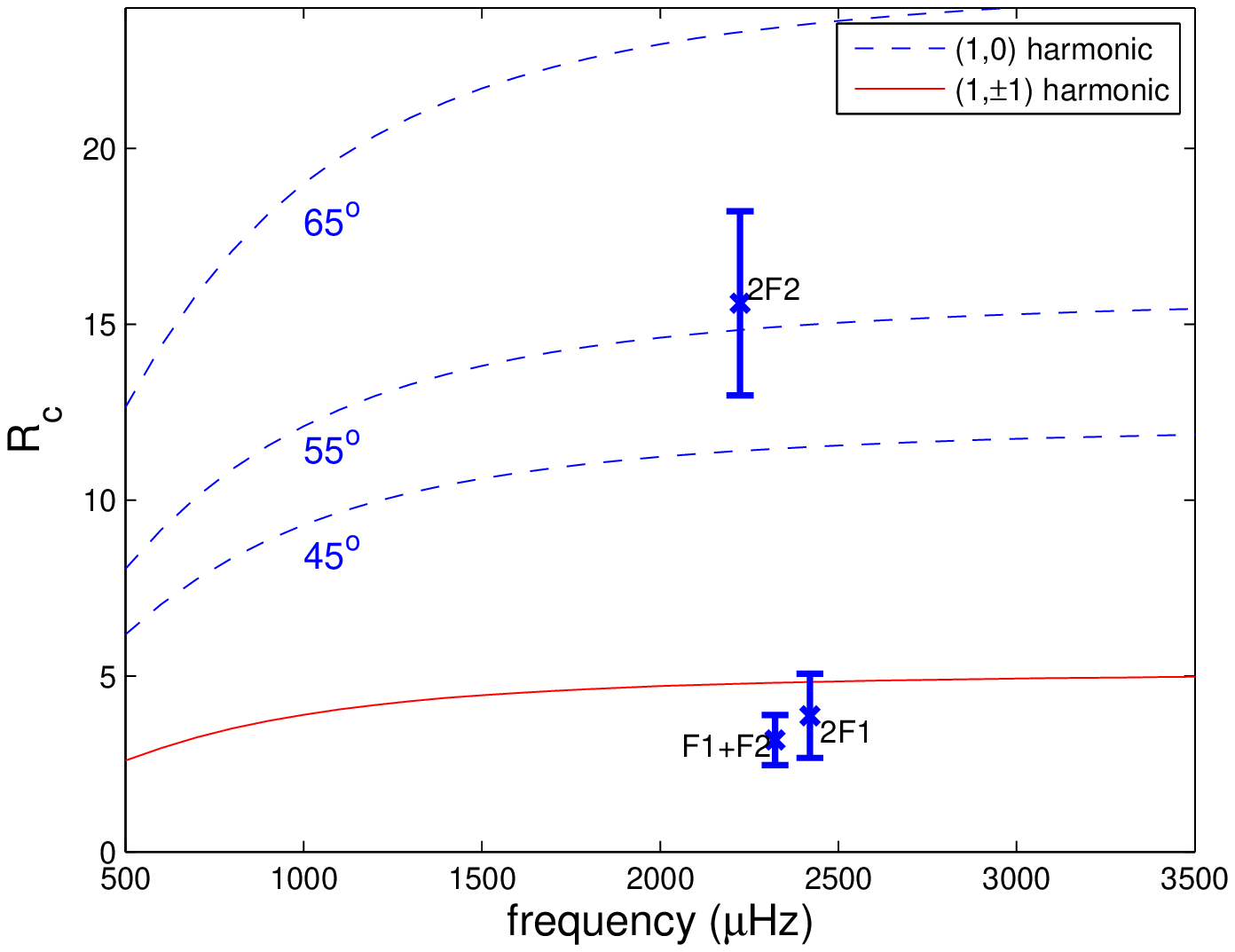}
\caption{\label{f:comb}The ratio between the combination mode and parent mode amplitudes (R$_c$) is plotted against the frequency of the combinations. Here we consider the harmonics of F1 and F2 along with their combination. The predicted values of \rc\ \citep{wu01, yeates05} for \el=1 modes are shown (blue as the \mzero\ harmonic (at inclinations of $45^\circ$, $55^\circ$ and 65$^\circ$) and red as the \mpm\ harmonic). The combination \rc\ values represent the average of the three photometric runs discussed in this paper, the error bar represents the scatter in those three measurements. The \el=2 combination modes have large \rc\ values, except for the harmonic of $(2,1)$ which lies near\rc=3.  This linear combination theory is consistent with the identification of F2 as \mzero\ and F1 as \mone.}
\end{figure}

Figure~\ref{f:comb} shows the predicted \rc\ values along with those values measured for the combinations of F1 and F2 on G29-38. We average the \rc\ values over the three photometric runs (September, October and November) and use a standard deviation of the three measurements to represent the error bars (the formal error bars are significantly smaller). The harmonic frequencies are most valuable in mode identification because the two parent modes have the same \lm. The harmonic 2F2 lies significantly higher than the harmonic 2F1.  This is only consistent with \citet{wu01} when F2 is an \mzero\ mode and F1 is \mone.   The combination theory then predicts the amplitude of a \mone\ combined with a \mzero\ to be the same amplitude as the harmonic of \mone \footnote{The \el=2 harmonics lie significantly off the plot in Figure~\ref{f:comb} for any inclination angle, except for $(2,1)$, which lies near R$_c\sim3$. However if F1 is $(2,1)$, the amplitude of F1+F2 should be much larger than observed.}. F1+F2 thereby supports the results of the harmonics. This theory clearly states F1=\mone\ and F2=\mzero, as suggested by the triplet structure. 

\subsection{Oblique Pulsator Model}
In certain types of pulsating stars, such as $\beta$ Cephei and roAp stars, the magnetic field can cause the pulsations to be aligned with an axis other than the rotation axis \citep{shibahashi00,kurtz86,kurtz82}. In this oblique pulsator mode, a single  pulsation will be split into $2\ell+1$ frequencies. These extra frequencies do not represent pulsation modes of the star, instead they are caused by a pulsation, not aligned with the rotation axis, rotating around the star.  If this is the cause for the triplets in our data of G29-38, the 15\,$\mu$Hz splittings of our two modes indicate a rotation period of $\sim$18.5\,hr.

\citet{montgomery10} offer the first evidence of oblique pulsation on a pulsating white dwarf star and so we mention this possibility for completeness.  If this model applies to G29-38, the asteroseismological arguments presented in \S\ref{s:trip} are not applicable. However, projecting the off-axis pulsation onto the rotation axis would yield triplets with exactly the same \lm\ geometry as determined from seismology, except that now the frequency splittings are given by the inverse of the rotation period. Thus, we arrive at the same mode identifications in either case.  

For a detailed analysis of the Ca geometry incorporating this oblique pulsator model, we would need to make refinements to the model presented by \citet{montgomery08}. However, our data are not sufficient to make any but the most general conclusions, so we leave this exercise to future papers should the oblique pulsator model be deemed appropriate for G29-38.

\section{Mapping Calcium Across G29-38's Surface}
\begin{figure*}[ht]
\centering
\includegraphics[scale=.55,angle=270]{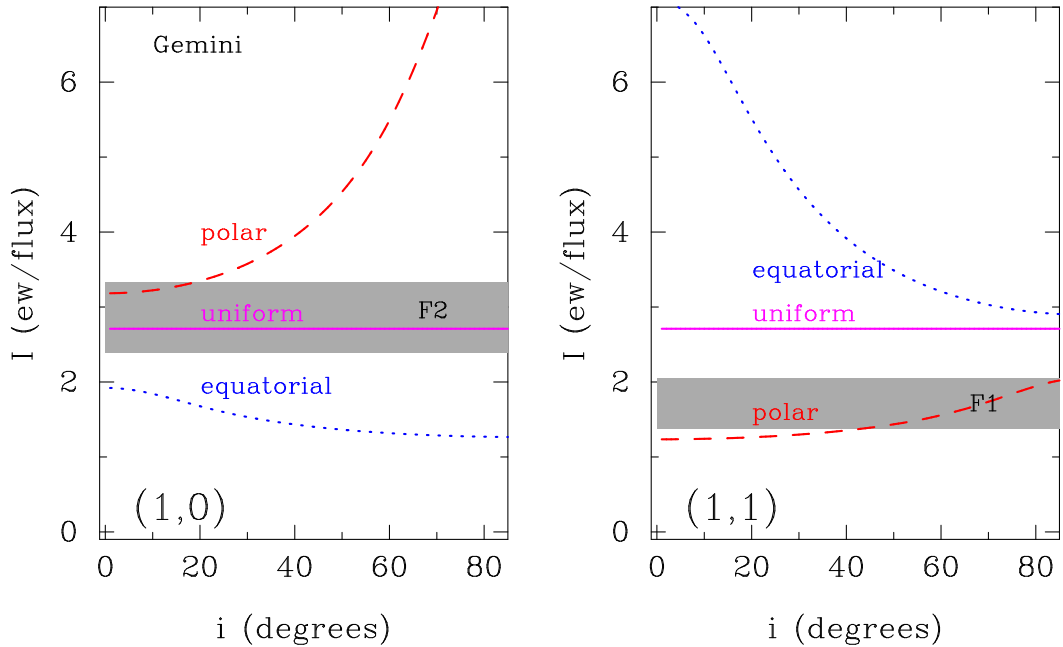}
\caption{\label{f:rvalues}
  The measured \ica\ values from the Gemini observations compared to the calculations of \citet{montgomery08} as a function of inclination angle ($i$).  The plot on the left depicts the models for the \mzero\ mode while the plot on the right depicts the \monep\ mode. The grey box represents the 1$\sigma$ error bars of the Gemini observations for F1 and F2. The observations favor the polar model.}
\end{figure*}

We compare our measurements of \ica$=A_{\rm EW}/A_{\rm flux}$ to the model presented by \citet{montgomery08} for F1 and F2; see Tables~5 and~6 for the \ica\ values for F1 and F2 in October and November. Because we see the same modes in both October and November, we take the average of the two measurements. F1, as a \mone\ mode, has an average \ica\ value of $1.7\pm0.3$, and F2, as a \mzero\ mode, has an average \ica\ value of $2.9\pm0.5$. Figure~\ref{f:rvalues} presents a direct comparison of these two measurements to the \citet{montgomery08} models (we note that the predicted \ica\ values for \monep\ are identical to those predicted for \mone). Next, we discuss which Ca geometry is most likely and then consider previous data sets from the VLT and Keck observatories.

\subsection{Comparison to the models}

First we consider the simplest case, a uniform accretion model. Here
the different geometries of the \m\ eigenvalues are inconsequential
because the Ca has a uniform geometry. All measurements of \ica\ for
the same \el\ should be equal. \citet{montgomery08} calculate that
\ica\ should be 2.7 for \el=1. Comparing our two measurements to this
model yields a $\chi^{2}_{r} = 4.3$, or less than a 1\% chance that
this is the appropriate model. Because of our inability to accurately
measure the continuum when measuring the EW, these two values may both
suffer from the same fractional offset in EW amplitude and thus \ica.
Even with the presence of such a systematic effect, the ratio of these
two measurements would be 1 if the uniform model is appropriate.
Instead, the ratio of their \ica\ values is $I_{\rm F1}/I_{\rm
  F2}=0.6\pm0.1$, indicating less than a 1\% chance these measurements
agree. Primarily, the uniform model fails to match F1 though it agrees
quite well with F2. With only one outlier from two measurements, our
data cannot completely rule out the uniform model, though they suggest
the surface is inhomogeneous.

Both the equatorial and polar models are more complex because the \ica\ values are dependent on the extent of the inhomogeneity, the inclination angle, and \lm. For F1 and F2, three separate techniques agree with the conclusion that F1 is \mone\ and F2 is \mzero. This leaves only the inclination angle and the extent of the inhomogeneity. \citet{montgomery05} has measured the inclination angle to be 65$^\circ$ by fitting the light curve of G29-38 and we find an inclination of 55$^\circ$ in \S\ref{s:comb} by using the amplitude of the harmonic of F2. However, we do not rely on these measurements for our results and only use the inclination measurements to suggest that the inclination angle is not at the extremes.  

We compare our observations to the model described by \citet{montgomery08}, where the rate of gravitational settling greatly exceeds the horizontal mixing rate by turbulent viscosity. However if the horizontal mixing rate were increased, or the location of the accretion is wider than assumed, the extent of the inhomogeneity for either the polar or equatorial model would decrease.  For the equatorial model we show in Figure~\ref{f:rvalues}, the ratio of these values, $I_{\rm (1,1)}/I_{\rm (1,0)}$, is between 2.3 and 3.7 depending on the inclination of the pulsation axis. The polar case is the exact opposite; $I_{\rm (1,0)}$ will always be larger than $I_{\rm (1,1)}$, with the ratio lying between 0.22 and 0.38. In both cases the ratio will approach one if the inhomogeneity is not as extreme as our model assumes.

Figure~\ref{f:rvalues} shows that the polar model fits better than the
equatorial model because $I_{\rm F1}/I_{\rm F2}=0.6\pm0.1<1$.  The
\ica\ values for F1 and F2 differ from the equatorial accretion model
by $\ge$11$\sigma$.  However, they differ from the polar accretion
model by 1.5--2.5$\sigma$, depending on the inclination angle.  In
both cases we could improve the fit by adjusting the amount of the
inhomogeneity, i.e., creating a larger region of Ca.  Only a minor
adjustment is needed to find a polar model that agrees with our
measurements. There is no way to adjust the models to find agreement
with the equatorial model.

While we cannot entirely eliminate the uniform distribution, based on
the two measurements of F1 and F2, we conclude that the Ca on G29-38
has a higher concentration at the poles than it does at the equator.

 
\subsection{Previous observations}
\label{s:pre}
Keck and VLT time series spectroscopy also exist for G29-38, primarily taken for the purpose of \el\ identification \citep{vk00,c00,thompson08}. For both, no adjoining photometry was taken, so all measurements of the pulsation flux amplitude are from the spectroscopy alone. While the photometry from the spectra is quite good, the frequency resolution is poor. The Keck and VLT spectral series are only 4 and 6\,hr in length, respectively. This does not allow us to distinguish splittings as close as the 15\,$\mu$Hz splittings that we see in the Gemini observations. As a result we cannot assume that our measurements are not influenced by unresolved modes. Keeping this in mind, we measure the \ica\ values of the largest modes in the 1996 Keck data and the 1999 VLT data in the manner explained above.

We use the reduced data sets of \citet{vk00} and \citet{thompson08} to measure the Ca EWs. The pulsation modes and the photometric amplitudes are taken directly from the respective papers. We measure the Ca EW for each individual spectrum as described in \S\ref{s:ew}. As with the Gemini data, we likely under-estimate the continuum, causing overly large values of Ca EW amplitudes. Using model white dwarf spectra emulating the VLT and Keck observations, we calculate the appropriate corrections as described in \S\ref{s:offset}. We can only measure the \ica\ values for the largest two modes in each data set, providing us four more measurements of the inhomogeneity of Ca on G29-38. Our average Ca EW for the Keck and VLT data after accounting for these offsets is $198\pm10$\,m\AA\ and $298\pm10$\,m\AA, respectively. The Keck data have a significantly lower Ca EW than the VLT or Gemini observations, a fact previously noted by \citet{vonhippel07}.

For these previous data sets to yield Ca inhomogeneity we need to know the \lm\ of the four modes. The line shape variations measured by \citet{thompson08} and \citet{c00} tell us that all 4 modes are $\ell=1$. The values for \m\ are less constrained.  We can use the combination frequencies, as we did in \S\ref{s:comb} to give us good guesses for \m.  Both the Keck and VLT data have a 615\,s mode. The amplitude of the harmonic for the 615\,s mode is low (\rc$\sim4$) in both the Keck and VLT data; by looking at Figure~\ref{f:comb}, this indicates that it is a \monep\ mode. \citet{montgomery05} also found this mode to be \monep\ by fitting G29-38's light curve. The other large amplitude Keck mode is at 817\,s, and the other VLT mode is at 810\,s. In both cases the amplitude of the harmonic is large ($R_{\rm c}\sim13$), like an \mzero\ harmonic should be. We tentatively conclude these modes are \mzero. See Table~\ref{t:Rother} for our EW and \ica\ measurements along with the best guess at \lm\ for the 4 modes we measured in the Keck and VLT data. 

\begin{figure}[t]
\includegraphics[scale=.35,angle=270]{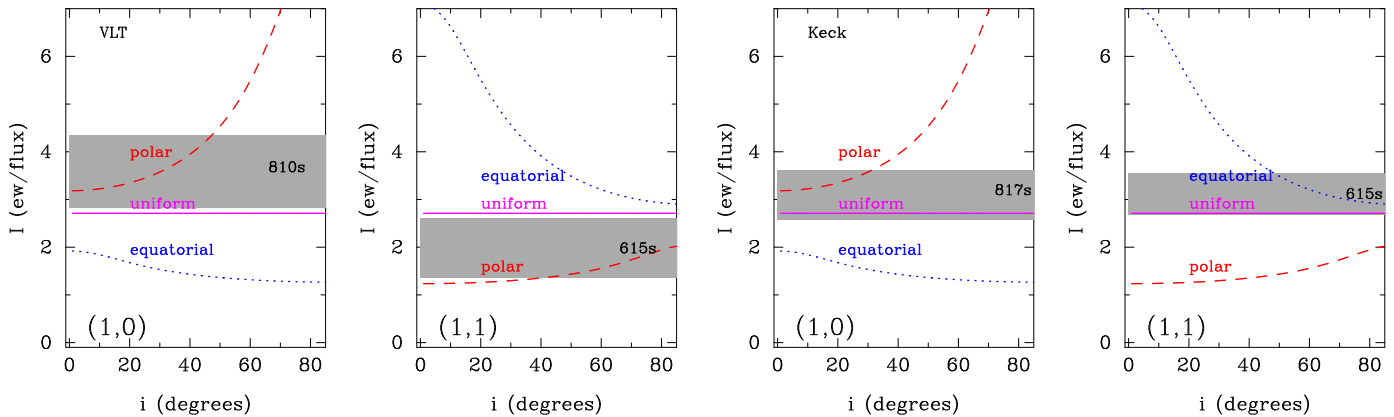}
\caption{\label{f:rvaluesVLT}The measured value for \ica\ using the 1999 VLT observations compared to the model values presented by \citet{montgomery08} as a function of inclination. The plot on the left depicts the models for the \mzero\ mode while the plot on the right depicts the \monep\ mode. The grey box represents the 1$\sigma$ error bars for the 615\,s and 810\,s modes. The measurements disagree with the equatorial model, see \S\ref{s:pre}.}
\end{figure}

Figures~\ref{f:rvaluesVLT} and \ref{f:rvaluesKeck} show a comparison of the VLT and Keck data to the polar, equatorial and uniform accretion models calculated by \citet{montgomery08}. The VLT values for \ica\ are very similar to the Gemini observations. The \monep\ mode is significantly smaller than the \mzero\ mode, $I_{\rm (1,1)}/I_{\rm (1,0)}=0.5\pm0.2$.  Again, this cannot rule-out the uniform distribution, but it disagrees with the equatorial model. The two Keck modes, however, have identical values for \ica. If these modes do indeed have different \m\ values, then the Keck data agrees best with a uniform distribution\footnote{\citet{montgomery08} used measurements from \citet{vonhippel07} that had not been corrected for offsets introduced by mis-measuring the continuum and thus had a much higher \ica\ value for the 615\,s mode (see \S\ref{s:offset}).}.  

Another indication that the 1996 inhomogeneity is different than the 1999 epoch is the 2$\sigma$ difference in the \ica\ measurement for the 615\,s, \monep\ mode.  A change in \ica\ indicates that the location of the Ca on the surface of G29-38 changed between 1996 and 1999. Except for the assumption that both data sets contain the same mode, this conclusion that the Ca geometry changed is independent of any model.

\begin{figure}[t]
\includegraphics[scale=.35,angle=270]{plots/RvaluesKeck.ps}
\caption{\label{f:rvaluesKeck}The measured value for \ica\ using the 1996 Keck observations compared to the model values presented by \citet{montgomery08} as a function of inclination. The plot on the left depicts the models for the \mzero\ mode while the plot on the right depicts the \monep\ mode. The grey box represents the 1$\sigma$ error bars for the 615\,s and 817\,s modes on the \monep\ and \mzero\ plots respectively. The data most closely match a uniform distribution of Ca, see \S\ref{s:pre}.}
\end{figure}

\subsection{Inhomogeneity from Velocities} 
\label{s:vel}
The g-mode pulsations cause the atmosphere to flow periodically across
the surface of the star, resulting in a periodic Doppler shift of the
Balmer lines at the period of the stellar pulsations. Due to the rapid
mixing by the turbulent motions which extend into the photosphere, the
pulsation velocities are independent of optical depth.  Also, the
amplitudes of the velocities depend on the geometry of pulsation, so
we can use them to find different values of \el\ \citep[see][]{vk00}.
In addition, the geometry of the velocity fields can indicate if any
surface inhomogeneity exists. For instance, if metals lie
inhomogeneously across the surface of G29-38, the amplitude of their
pulsational velocities would not necessarily equal those measured from
the H lines. The geometry of the velocity variations for these g-modes is proportional to the derivative of the flux variations.  Thus, a larger velocity
amplitude from the metal line compared to the H lines could result if
the metals were concentrated where the velocities are the largest, near the flux nodal line, e.g., the equator of an \mzero\ mode.


We fit the H$\beta$ and H$\gamma$ lines as described in \citet{thompson04}, with a combination of a Lorentzian and a Gaussian to measure the periodic Doppler shifts in the Gemini spectra.  The Ca~K line is much smaller and we only fit a Gaussian to measure its velocity.  We then create velocity curves for both the average H and the Ca~K lines.  Similarly, we measure the Ca~K lines for both the Keck and VLT data sets; their H line amplitudes have already been published by \citet{vk00} and \citet{thompson08}.  

We measure the velocity at F1 and F2 for the H velocity curves from the Gemini data (see Tables 5 and 6). For the H line, we also measure the ratio between the velocity and flux amplitudes ($R_{\rm v}=(A_{\rm v}/2\pi f)/A_{\rm flux}$).  Our velocity amplitudes agree with previous measurements of \el=1 modes on G29-38, further evidence that F1 and F2 are \el=1 for our modes.  
We were unfortunately unable to measure the velocity amplitudes of the Ca~K line from the 2008 Gemini and the 1999 VLT data.  Both show no significant peaks in the FT with an uninteresting limit of 11\,\kms.  This is significantly larger than the H velocity amplitude and thus we can conclude nothing about the Ca inhomogeneity for these two data sets from the velocities alone.

However, the 1996 Keck Ca velocity FT shows a significant peak for the 615\,s mode with a velocity of 24$\pm6$\,\kms.  This measured velocity amplitude is 4.4$\pm1.3$ times larger than the H velocity amplitude (5.4$\pm$0.8\,\kms) measured by \citet{vk00} for the same mode. It is also worth noting that another peak is present in the Ca velocity FT, at 7505\,$\mu$Hz with an amplitude of 22$\pm6$\,\kms.  This 133\,s mode does not appear in the Keck photometric FT and may indicate that these Ca velocity measurements are actually dominated by an unknown systematic effect.  If the 615\,s Ca velocity peak is real, the larger Ca velocity amplitude indicates that Ca is located on the surface in a way that maximizes the line-of-sight velocities. 

Precise modeling of the velocities that include the inclination angle is needed to see how the Ca could be distributed in order for its velocity amplitude to be much larger than the H velocity amplitude. As an example of how this might occur, the 615\,s mode is \monep\ and the largest line-of-sight velocities will come from the limb of the star for high inclinations. If all the visible Ca were located in the limb of the star during the Keck observations, only the largest, line-of-sight velocities would be observed for Ca, while the H line would include regions with no line-of-sight velocity shifts.  At the very least, this difference in velocity amplitude is an indication that Ca is inhomogeneous in 1996, when the Keck observations were obtained. However the difference in velocities contradicts the \ica\ measurements for the Keck data which favor a uniform distribution of Ca.  Possibly more complicated scenarios including spots of Ca could reconcile these two measurements.

%

\section{Discussion of Accretion}
If we consider our results for each epoch of measurement (1996, 1999 and 2008), then something about the accretion has changed in 1996 (Keck) when compared to 1999 (VLT) and 2008 (Gemini). The 1999 and 2008 results indicate that the majority of the Ca accretes at the poles.  This is most likely a signpost of magnetically-influenced accretion.  However, in 1996 the accretion appears to be different; the \ica\ values indicate a uniform accretion.  Perhaps the unusual results from the 1996 epoch should not be a surprise as \citet{vonhippel07} already noted a change in accretion at this time. The Ca EW was 70\% smaller than in 1999 and the Ca EW showed fluctuations on time scales as short as 2 weeks \citep[but also see][]{debes08}. Once again, the Keck data show that the accretion of G29-38 was not entirely constant. 

These variations could be used to better understand either the disk or the accretion process. Since the accretion geometry indicates that the magnetic field was not influencing accretion in 1996, exactly when the EW was smallest, any Ca EW modulation may be connected to a modulation in the magnetic field.

We do not know the magnetic field strength at the surface of G29-38.
The non-detection of Zeeman splitting in this star only limits the
magnetic field to be $\la$ 10\,kG \citep{koester98}. Yet the apparent
detection of polar accretion implies magnetic fields. If G29-38's
magnetic field is strong enough and has a complex, evolving structure,
then this magnetic field could modulate the quantity of gas that is
accreted onto the star. In this case, the variations in surface Ca
abundance would indicate that the timescale for significant changes in
the magnetic field structure is less than two weeks.  The plausibility
of this scenario is, at present, limited by our knowledge of the field
strength, how it influences accretion rates, and the physical
mechanisms that would cause G29-38's magnetic field to change.

This time scale of two weeks places constraints on the magnitude and
global geometry of the proposed field. If the field has a large scale
structure (say dipolar) and is non-zero outside of the star, then the
energy contained in such a field is approximately
\begin{equation}
  E \approx \langle B^2 \rangle V,
\end{equation}
where $V$ is the volume of the star and $\langle B^2 \rangle$ is the
value of $B^2$ averaged over the stellar surface. For a 10~kG field
this is $\sim 10^{35}$~ergs. This is much larger than the total energy
radiated into space by the star in a two week period, implying that
building such a field on this time scale is not possible. On the other
hand, if the field is less global and is non-zero \emph{only} in the
surface layers, then the energy in a 10~kG field would require only
about 1\% of the energy radiated by the star in a two week period,
which is energetically feasible. Thus, we may be seeing the effect of
smaller-scale, local variations in the magnetic field which
nevertheless are able to perturb the accretion geometry.

\section{Conclusions}
We combined our Gemini spectroscopy and Whole Earth Telescope photometry to constrain the accretion geometry of Ca on G29-38. First, we identified the pulsation modes using a variety of techniques that took advantage of all the information contained in the time-series spectroscopy and photometry. Second, we measured the Ca EW pulsation amplitudes and compared them to the flux amplitudes. This combination of measurements allowed us to constrain the geometry of Ca on the surface of G29-38. These data from 2008 somewhat suggest the Ca is inhomogeneous: a polar distribution of Ca is most likely and the equatorial distribution has been ruled out.  If the Ca does lie at or near the poles, the accretion must be occurring there. This is the first evidence that a magnetic field is influencing the accretion process onto G29-38. 

We have shown how this technique may be used to directly measure the distribution of Ca on the surface of a white dwarf and thereby infer the accretion geometry; however a few improvements would have made our results more conclusive.  Since we were allocated band-3 Gemini time, we chose a lower resolution grating to increase our odds of getting Gemini observations. A higher resolution would have improved our Gaussian fits of the Ca line and also helped establish the continuum across the line.  Our photometry was significantly better in September, and had our spectroscopy come at the same time, we would have reduced our flux amplitude error bars by 30\%.  The best way to improve the strength of our conclusions is to simply take more measurements. With approximately twice as much time series spectroscopy we would have reduced the error bars by 40\% and then we could have convincingly distinguished between a polar and a uniform model. 

While previous time-series spectroscopy on G29-38 did not have the frequency resolution of our data, these data both confirm our polar conclusions (VLT observations) and present the tantalizing possibility that the accretion geometry may change (Keck observations) over time. As the Keck observations also coincide with a lower apparent Ca abundance, we postulate how the accretion geometry and rate may be related. With further observations of G29-38 we may establish that accretion in this white dwarf debris disk system is not only influenced by magnetic fields but is also dynamic in nature.

\acknowledgements SET and MHM thank the Crystal Trust for the
financial support. TvH and MHM thank the National Science Foundation
for support under grants AST 06-07480 and AST-0909107, respectively.
We thank all the observers of the Whole Earth Telescope for being
flexible with their observing schedules. PROMPT observations were made possible by the Robert Martin Ayers Science Fund. This work is based on observations obtained at the Gemini Observatory (ID: GN-2008B-Q-124),
which is operated by the Association of Universities for Research in
Astronomy, Inc., under a cooperative agreement with the NSF on behalf
of the Gemini partnership: the National Science Foundation (United
States), the Science and Technology Facilities Council (United
Kingdom), the National Research Council (Canada), CONICYT (Chile), the
Australian Research Council (Australia), Minist\'erio da Ci\^encia e
Tecnologia (Brazil) and Ministerio de Ciencia, Tecnolog\'ia e
Innovaci\'on Productiva (Argentina)

\bibliography{davzds}
\bibliographystyle{apj}

\input{plots/reportsept.tab}
\input{plots/reportoct.tab}
\input{plots/reportnov.tab}

\input{plots/sept.tab} 
\input{plots/gemn1.tab}

\input{plots/gemn2.tab}

\input{plots/Rother.tab}

\end{document}

%% file: plots/reportsept.tab
\begin{table}[ht]
\centering
\caption{Photometric observations of G29-38 in September 2008.}
\begin{tabular}{llll}
Observatory & Date &  Start(UTC) & Length (hr)\\
\hline
\hline
OPD Brazil & 2008-09-20 & 03:55:17 & 3:12:21\\
Apache Point 1m & 2008-09-21 & 05:53:39 & 5:05:41\\
Meyer Observatory 0.6m  & 2008-09-21 & 06:17:08 & 4:14:50\\
Meyer Observatory 0.6m  & 2008-09-22 & 03:01:27 & 6:59:50\\
OPD Brazil & 2008-09-23 & 00:37:13 & 5:32:29\\
McDonald Observatory 2.1m & 2008-09-23 & 02:11:03 & 9:27:25\\
Meyer Observatory 0.6m  & 2008-09-23 & 02:23:40 & 4:40:39\\
Apache Point 1m & 2008-09-23 & 04:50:44 & 6:04:59\\
OPD Brazil & 2008-09-24 & 00:18:02 & 6:56:20\\
McDonald Observatory 2.1m & 2008-09-24 & 03:10:58 & 2:00:25\\
Meyer Observatory 0.6m  & 2008-09-24 & 03:13:03 & 7:29:50\\
Apache Point 1m & 2008-09-24 & 07:47:40 & 2:28:42\\
McDonald Observatory 2.1m & 2008-09-24 & 08:13:13 & 2:51:20\\
McDonald Observatory 2.1m & 2008-09-25 & 02:06:36 & 6:07:00\\
Meyer Observatory 0.6m  & 2008-09-25 & 04:13:00 & 5:59:50\\
Mcdonald Observatory 2.1m & 2008-09-26 & 02:04:08 & 8:14:15\\
Meyer Observatory 0.6m  & 2008-09-26 & 06:27:05 & 3:59:49\\
Apache Point 1m & 2008-09-26 & 07:42:35 & 3:03:31\\
Tuebingen Observatory & 2008-09-26 & 19:12:43 & 7:18:46\\
McDonald Observatory 2.1m & 2008-09-27 & 02:02:32 & 7:22:05\\
Meyer Observatory 0.6m  & 2008-09-27 & 02:02:55 & 7:59:50\\
Mt. Suhora 0.6m & 2008-09-27 & 18:37:59 & 5:35:00\\
Krakow 0.5m & 2008-09-27 & 18:53:04 & 2:17:10\\
Tuebingen Observatory & 2008-09-27 & 20:30:24 & 6:25:36\\
McDonald Observatory 2.1m & 2008-09-28 & 02:06:17 & 9:07:10\\
Krakow 0.5m & 2008-09-28 & 17:33:49 & 7:06:41\\
McDonald Observatory 2.1m & 2008-09-29 & 02:57:47 & 7:51:55\\
Meyer Observatory 0.6m  & 2008-09-29 & 07:31:01 & 2:59:50\\
Meyer Observatory 0.6m  & 2008-09-30 & 04:19:31 & 4:59:50\\
\hline
\end{tabular}
\end{table}	

%% file: plots/reportoct.tab
\begin{table}[h]
\centering
\caption{Photometric observations of G29-38 in October 2008.  }
\begin{tabular}{llll}
Observatory & Date &  Start(UTC) & Length (hr)\\
\hline
\hline
Krakow 0.5m & 2008-10-05 & 21:11:58 & 2:10:43\\
Mt. Cuba 0.6m & 2008-10-08 & 01:51:51 & 4:05:58\\
Krakow 0.5m & 2008-10-08 & 18:35:56 & 3:23:58\\
Krakow 0.5m & 2008-10-11 & 18:00:54 & 5:44:32\\
Apache Point 1m & 2008-10-17 & 01:23:51 & 7:50:31\\
Mt. Cuba 0.6m & 2008-10-17 & 03:50:02 & 2:50:14\\
Mt. Suhora 0.6m & 2008-10-17 & 17:29:49 & 1:18:12\\
Apache Point 1m & 2008-10-18 & 02:36:09 & 6:32:30\\
Gemini North 8m & 2008-10-18 & 06:25:34 & 2:17:02\\
Krakow 0.5m & 2008-10-18 & 16:45:29 & 7:06:00\\
Krakow 0.5m & 2008-10-19 & 18:36:40 & 1:06:00\\
Mt. Cuba 0.6m & 2008-10-20 & 00:03:13 & 3:58:43\\
Mt. Cuba 0.6m & 2008-10-21 & 01:41:57 & 4:40:41\\
\hline
\end{tabular}
\end{table}	

%% file: plots/reportnov.tab
\begin{table}[h]
\centering
\caption{Photometric observations of G29-38 in November 2008.  }
\begin{tabular}{llll}
Observatory & Date &  Start(UTC) & Length (hr)\\
\hline
\hline
SOAR 4m & 2008-11-02 & 03:55:00 & 1:51:08\\
PROMPT 0.4m & 2008-11-03 & 02:23:29 & 3:17:50\\
Vienna 0.8m & 2008-11-03 & 16:50:11 & 6:28:23\\
Mt. Suhora 0.6m & 2008-11-03 & 18:57:20 & 5:33:02\\
Apache Point 1m & 2008-11-04 & 01:29:28 & 6:02:59\\
PROMPT 0.4m & 2008-11-04 & 02:25:29 & 2:36:10\\
Mt. Suhora 0.6m & 2008-11-04 & 17:19:39 & 2:40:02\\
PROMPT 0.4m & 2008-11-05 & 01:37:41 & 4:13:26\\
Mt. Suhora 0.6m & 2008-11-05 & 17:18:47 & 6:02:12\\
Apache Point 1m & 2008-11-06 & 01:55:29 & 5:38:22\\
Apache Point 1m & 2008-11-07 & 01:35:45 & 5:27:49\\
Apache Point 1m & 2008-11-11 & 02:53:48 & 4:37:43\\
Krakow 0.5m & 2008-11-11 & 16:50:38 & 5:44:40\\
Apache Point 1m & 2008-11-16 & 03:10:10 & 3:47:56\\
Apache Point 1m & 2008-11-17 & 00:58:56 & 5:53:54\\
\hline
\end{tabular}
\end{table}

%% file: plots/sept.tab
\begin{table}[h]
\label{t:sept}
\centering
\caption{Measured variations in brightness of G29-38 from September 2008. }
\begin{tabular}{rllll}
\hline
mode & frequency & A$_{\rm flux}$ & T$_{\rm 0,flux}$ & R$_{\rm c}$ \\
  & $\mu$Hz & \%  & s &  \\

\hline
\hline
F1 & 1209.610$\pm$0.006 & 3.04$\pm$0.02 & 742$\pm$1 & ------\\
F2 & 1112.443$\pm$0.007 & 2.15$\pm$0.03 & 779$\pm$2 & ------\\
F3 & 1224.701$\pm$0.009 & 1.91$\pm$0.02 & 579$\pm$2 & ------\\
F4 & 1239.721$\pm$0.014 & 1.32$\pm$0.03 & 654$\pm$2 & ------\\
F5 & 1127.520$\pm$0.016 & 1.17$\pm$0.02 & 142$\pm$4 & ------\\
F6 & 1097.446$\pm$0.017 & 1.03$\pm$0.03 & 791$\pm$4 & ------\\
F7 & 1293.038$\pm$0.023 & 0.78$\pm$0.02 & 561$\pm$4 & ------\\
F8 & 1522.498$\pm$0.019 & 0.93$\pm$0.03 & 144$\pm$3 & ------\\
F9 & 2502.323$\pm$0.013 & 1.31$\pm$0.03 & 291$\pm$1 & ------\\
F10 & 1999.767$\pm$0.028 & 0.63$\pm$0.02 & 154$\pm$3 & ------\\
F11 & 1015.214$\pm$0.031 & 0.59$\pm$0.36 & 890$\pm$291 & ------\\
=2F1 & 2419.220$\pm$0.000 & 0.41$\pm$0.03 & 412$\pm$4 & 4.4$\pm$0.3\\
=2F2 & 2224.886$\pm$0.000 & 0.76$\pm$0.02 & 38$\pm$2 & 16.3$\pm$0.6\\
=F1+F2 & 2322.053$\pm$0.000 & 0.35$\pm$0.02 & 422$\pm$5 & 2.7$\pm$0.1\\
=F1+F3 & 2434.311$\pm$0.000 & 0.24$\pm$0.02 & 350$\pm$7 & 2.1$\pm$0.1\\
=F2+F3 & 2337.144$\pm$0.000 & 0.50$\pm$0.03 & 383$\pm$4 & 6.1$\pm$0.2\\
=F2+F4 & 2352.164$\pm$0.000 & 0.34$\pm$0.02 & 368$\pm$5 & 6.0$\pm$0.2\\
=2F4 & 2479.442$\pm$0.000 & 0.13$\pm$0.02 & 75$\pm$16 & 7.4$\pm$1.5\\
=F3+F4 & 2464.422$\pm$0.000 & 0.16$\pm$0.02 & 253$\pm$11 & 3.1$\pm$0.2\\
=F1+F6 & 2307.056$\pm$0.000 & 0.19$\pm$0.03 & 9$\pm$9 & 3.0$\pm$0.2\\
\hline
\end{tabular}
\\
Note: Variations are measured using the form A$\sin(2\pi f(t+T_0))$. All T$_{0}$ values are measured from BJED=2454700. R$_{\rm c}$ is the ratio of the parent to harmonic amplitudes defined in \S\ref{s:comb}.
\end{table}

%% file: plots/gemn1.tab
\begin{sidewaystable}[]
\centering
\caption{Measured variations in brightness, velocity and Ca EW in October 2008.}
\begin{tabular}{rllll|lll|lll}
\hline
mode & frequency & A$_{\rm flux}$ & T$_{\rm 0,flux}$ & R$_{\rm c}$ & A$_{\rm EW}$ & T$_{\rm 0,EW}$ & I$_{\rm Ca}$ &
A$_{\rm vel}$ & T$_{\rm 0,vel}$ & R$_{\rm v}$ \\
    & $\mu$Hz & \%  & s & &\% & s  &  &  km\,s$^{-1}$ & s & ${\rm Mm\, rad}^{-1}$\\
\hline
\hline
F1 & 1210.642$\pm$0.005 & 3.48$\pm$0.04 & 755$\pm$1 & ------ & 6.8$\pm$1.9 & 150$\pm$37 & 1.96$\pm$0.56 & 3.6$\pm$0.6 & 97$\pm$21 & 13.7$\pm$2.2\\
F2 & 1111.806$\pm$0.005 & 2.77$\pm$0.04 & 310$\pm$2 & ------ & 8.4$\pm$1.9 & 68$\pm$33 & 3.03$\pm$0.70 & 2.6$\pm$0.6 & 451$\pm$32 & 13.6$\pm$3.0\\
F3 & 1224.716$\pm$0.011 & 1.44$\pm$0.04 & 664$\pm$3 & ------ & ----- & ----- & ----- & ----- & ----- & -----\\
F4 & 1238.990$\pm$0.021 & 0.66$\pm$0.04 & 279$\pm$7 & ------ & ----- & ----- & ----- & ----- & ----- & -----\\
F5 & 1125.730$\pm$0.023 & 0.68$\pm$0.04 & 151$\pm$8 & ------ & ----- & ----- & ----- & ----- & ----- & -----\\
F6 & 1098.795$\pm$0.015 & 1.04$\pm$0.04 & 217$\pm$5 & ------ & ----- & ----- & ----- & ----- & ----- & -----\\
F7 & 1292.522$\pm$0.046 & 0.34$\pm$0.04 & 129$\pm$13 & ------ & ----- & ----- & ----- & ----- & ----- & -----\\
F8 & 1522.618$\pm$0.016 & 0.94$\pm$0.04 & 461$\pm$4 & ------ & ----- & ----- & ----- & ----- & ----- & -----\\
F9 & 2502.286$\pm$0.014 & 1.10$\pm$0.04 & 249$\pm$2 & ------ & ----- & ----- & ----- & ----- & ----- & -----\\
F10 & 1999.676$\pm$0.015 & 1.01$\pm$0.04 & 116$\pm$3 & ------ & ----- & ----- & ----- & ----- & ----- & -----\\
F11 & 1011.818$\pm$0.021 & 0.75$\pm$0.04 & 294$\pm$8 & ------ & ----- & ----- & ----- & ----- & ----- & -----\\
=2F1 & 2421.284$\pm$0.000 & 0.30$\pm$0.04 & 13$\pm$8 & 2.5$\pm$0.3 & ----- & ----- & ----- & ----- & ----- & -----\\
=2F2 & 2223.613$\pm$0.000 & 0.97$\pm$0.04 & 24$\pm$3 & 12.7$\pm$0.5 & ----- & ----- & ----- & ----- & ----- & -----\\
=F1+F2 & 2322.448$\pm$0.000 & 0.55$\pm$0.04 & 223$\pm$5 & 2.9$\pm$0.1 & ----- & ----- & ----- & ----- & ----- & -----\\
=F1+F3 & 2435.358$\pm$0.000 & 0.20$\pm$0.04 & 373$\pm$12 & 2.0$\pm$0.2 & ----- & ----- & ----- & ----- & ----- & -----\\
=F2+F3 & 2336.523$\pm$0.000 & 0.62$\pm$0.04 & 168$\pm$4 & 7.9$\pm$0.3 & ----- & ----- & ----- & ----- & ----- & -----\\
=F2+F4 & 2350.797$\pm$0.000 & 0.20$\pm$0.04 & 413$\pm$13 & 5.5$\pm$0.5 & ----- & ----- & ----- & ----- & ----- & -----\\
=2F4 & 2477.981$\pm$0.000 & 0.13$\pm$0.04 & 335$\pm$19 & 29.9$\pm$9.0 & ----- & ----- & ----- & ----- & ----- & -----\\
=F3+F4 & 2463.707$\pm$0.000 & 0.21$\pm$0.04 & 119$\pm$12 & 10.9$\pm$1.1 & ----- & ----- & ----- & ----- & ----- & -----\\
=F1+F6 & 2309.437$\pm$0.000 & 0.35$\pm$0.04 & 200$\pm$7 & 4.8$\pm$0.3 & ----- & ----- & ----- & ----- & ----- & -----\\
\hline
\end{tabular}
\\
Note: Variations are measured using the form A$\sin(2\pi f(t+T_0))$. All T$_{0}$ values are measured from BJED=2454700.
\end{sidewaystable}

%% file: plots/gemn2.tab
\begin{sidewaystable}[]
\centering
\caption{Measured variations in brightness, velocity and Ca EW in November 2008.}
\begin{tabular}{rllll|lll|lll}
\hline
mode & frequency & A$_{\rm flux}$ & T$_{\rm 0,flux}$ & R$_{\rm c}$ & A$_{\rm EW}$ & T$_{\rm 0,EW}$ & I$_{\rm Ca}$ &
A$_{\rm vel}$ & T$_{\rm 0,vel}$ & R$_{\rm v}$ \\
    & $\mu$Hz & \%  & s & &\% & s  &  &  km\,s$^{-1}$ & s & ${\rm Mm\, rad}^{-1}$\\
\hline
\hline
F1 & 1210.710$\pm$0.006 & 3.73$\pm$0.03 & 21$\pm$1 & ------ & 5.5$\pm$1.6 & 334$\pm$38 & 1.47$\pm$0.43 & 3.2$\pm$0.7 & 127$\pm$27 & 11.3$\pm$2.3\\
F2 & 1112.606$\pm$0.008 & 2.46$\pm$0.03 & 192$\pm$2 & ------ & 6.6$\pm$1.6 & 757$\pm$34 & 2.68$\pm$0.66 & 3.7$\pm$0.7 & 381$\pm$25 & 21.6$\pm$3.8\\
F3 & 1224.899$\pm$0.022 & 0.93$\pm$0.03 & 73$\pm$4 & ------ & ----- & ----- & ----- & ----- & ----- & -----\\
F4 & 1237.013$\pm$0.148 & 0.14$\pm$0.03 & 207$\pm$29 & ------ & ----- & ----- & ----- & ----- & ----- & -----\\
F5 & 1126.067$\pm$0.035 & 0.59$\pm$0.03 & 25$\pm$8 & ------ & ----- & ----- & ----- & ----- & ----- & -----\\
F6 & 1097.029$\pm$0.044 & 0.47$\pm$0.03 & 556$\pm$10 & ------ & ----- & ----- & ----- & ----- & ----- & -----\\
F7 & 1293.292$\pm$0.105 & 0.20$\pm$0.03 & 262$\pm$20 & ------ & ----- & ----- & ----- & ----- & ----- & -----\\
F8 & 1522.459$\pm$0.035 & 0.59$\pm$0.03 & 482$\pm$6 & ------ & ----- & ----- & ----- & ----- & ----- & -----\\
F9 & 2502.317$\pm$0.018 & 1.15$\pm$0.03 & 322$\pm$2 & ------ & ----- & ----- & ----- & ----- & ----- & -----\\
F10 & 1999.583$\pm$0.021 & 1.00$\pm$0.03 & 400$\pm$3 & ------ & ----- & ----- & ----- & ----- & ----- & -----\\
F11 & 1014.329$\pm$0.020 & 1.04$\pm$0.03 & 605$\pm$5 & ------ & ----- & ----- & ----- & ----- & ----- & -----\\
=2F1 & 2421.420$\pm$0.032 & 0.65$\pm$0.03 & 80$\pm$3 & 4.7$\pm$0.2 & ----- & ----- & ----- & ----- & ----- & -----\\
=2F2 & 2225.212$\pm$0.019 & 1.08$\pm$0.03 & 388$\pm$2 & 17.8$\pm$0.6 & ----- & ----- & ----- & ----- & ----- & -----\\
=F1+F2 & 2323.316$\pm$0.028 & 0.73$\pm$0.03 & 188$\pm$3 & 4.0$\pm$0.1 & ----- & ----- & ----- & ----- & ----- & -----\\
=F1+F3 & 2435.609$\pm$0.194 & 0.11$\pm$0.03 & 166$\pm$19 & 1.5$\pm$0.2 & ----- & ----- & ----- & ----- & ----- & -----\\
=F2+F4 & 2349.619$\pm$0.079 & 0.26$\pm$0.03 & 285$\pm$8 & 38.4$\pm$4.9 & ----- & ----- & ----- & ----- & ----- & -----\\
=2F4 & 2474.026$\pm$0.091 & 0.23$\pm$0.03 & 167$\pm$9 & 1166$\pm$409 & ----- & ----- & ----- & ----- & ----- & -----\\
=F3+F4 & 2461.912$\pm$0.162 & 0.13$\pm$0.03 & 328$\pm$16 & 49.4$\pm$8.4 & ----- & ----- & ----- & ----- & ----- & -----\\
=F1+F6 & 2307.739$\pm$0.092 & 0.23$\pm$0.03 & 417$\pm$10 & 6.4$\pm$0.5 & ----- & ----- & ----- & ----- & ----- & -----\\
\hline
\end{tabular}
\\
Note: Variations are measured using the form A$\sin(2\pi f(t+T_0))$. All T$_{0}$ values are measured from BJED=2454700.
\end{sidewaystable}

%% file: plots/Rother.tab
\begin{table}[h]
\centering
\caption{Measured Ca EW and R values for the Keck and VLT observations.  }
\label{t:Rother}
\begin{tabular}{lllll}
mode & Period (s)  & A$_{\rm EW}$ (\%) & \ica & \lm \\
\hline
\hline
Keck\,1 & 614.20 & 9.9$\pm$1.4 & 3.1$\pm$0.4 & \monep\\
Keck\,2 & 817.66 & 8.4$\pm$1.4 & 3.1$\pm$0.5 & \mzero\\
VLT\,1 & 615.38 & 5.5$\pm$1.7 & 2.0$\pm$0.6 & \monep\\
VLT\,2 & 810.76 & 8.2$\pm$1.7 & 3.6$\pm$0.8 & \mzero\\
\hline
\end{tabular}
\end{table}